%% file: main.tex
\documentclass[10pt,twocolumn]{article}
\usepackage[letterpaper,margin=1in]{geometry}
\usepackage{graphicx}
\usepackage{amsmath,amssymb,amsfonts}
\usepackage{amsthm}
\usepackage{tikz}
\usepackage[version=4]{mhchem}
\usepackage{url}
\usepackage{makecell}
\usepackage[switch]{lineno} 

\usepackage{caption}
\usepackage{algorithm}
\usepackage{algorithmic}
\usepackage[sort&compress,numbers]{natbib}
\usepackage{hyperref}
\usepackage[export]{adjustbox}

\newcommand{\TotalCrystalStructures}{81 million}

\newcommand{\StableProtostructuresThatDoNotExitInAlexandria}{2303} 
\newcommand{\PrototypesUnderHundred}{91,295} 

\newcommand{\PrototypesUnderHundredNovel}{88,498} 
\newcommand{\ProtoStructuresUnderHundred}{456,110} 

\newcommand{\NumberOfPhaseDiagrams}{4495}

\title{Screening 39 billion protostructures for materials discovery}

\author{Abhijith~S~Parackal$^{1}$, Florian~Trybel$^{1}$, Felix~Andreas~Faber$^{2,\dagger}$, Rickard~Armiento$^{1,*}$\\
\small $^{1}$Department of Physics, Chemistry and Biology, Linköping University, Linköping, SE-581 83, Sweden\\
\small $^{2}$ Predictive Science, Digital and Automation, Pharmaceutical Sciences R$\&$D, AstraZeneca, Gothenburg, Sweden\\
\small $^{\dagger}$Corresponding author. Email: felix.faber@astrazeneca.com
\small $^{*}$Corresponding author. Email: rickard.armiento@liu.se}

\date{}

\begin{document}
\include{alias}

\twocolumn[{%
 \centering
\maketitle

\renewenvironment{abstract}{\small\quotation\noindent\textbf{Abstract:}}{\endquotation}

\begin{abstract}
Large-scale computational surveys are increasingly used to map the landscape of stable crystalline materials. We report a high-throughput energy screening of inorganic crystals that enumerates binary and ternary compositions up to a specified unit-cell complexity, yielding 39 billion protostructures. Candidates predicted to lie on or near the convex hull are retained, and their degrees of freedom are explored via Latin hypercube sampling followed by relaxation with machine-learned interatomic potentials. The resulting dataset contains \TotalCrystalStructures\ locally relaxed crystal structures spanning \NumberOfPhaseDiagrams\ ternary phase diagrams 
constructed from elements ranging from lithium to bromine 
and contains \PrototypesUnderHundredNovel\ crystal prototypes not present in existing crystal-structure databases. The methods are validated both for three well-explored materials systems, Zr-Zn-N, Ti-Zn-N, and Hf-Zn-N, and by comparing with known data for structures resulting from the larger screening. The work provides a systematic map of low-energy compositional–structural space and a large, structured pool of candidates for downstream property evaluation and materials design. 
\end{abstract}


\noindent\rule{\linewidth}{0.4pt}

\strut

}]

\noindent The field of crystal structure prediction (CSP) is central to materials science and drives materials discovery \cite{Oganov2019}. Recent advances in machine learning (ML) are transforming this field.
Large-scale models, such as deep graph networks trained on huge amounts of materials data, predict crystal stability with impressive accuracy and generalization~\cite{Merchant2023, Schmidt2023} and are coupled with various techniques for optimization and exploration. However, even with these tools, the full space of all possible materials appears too large for systematic, unbiased exploration \cite{walshenum, Hornfeck:uv5014}.

Most ML models and the associated approaches to the CSP problem rely on data from a range of well-known large open repositories of crystal structures, such as AFLOW~\cite{AFLOWLibraryCrystallographic}, Materials Project~\cite{jainCommentaryMaterialsProject2013}, NOMAD~\cite{Scheidgen2023}, OQMD~\cite{kirklin_open_2015}, and Alexandria~\cite{Schmidt2023,sciadvabi7948,schmidtDataset175kStable2022,schmidt2024}. 
Their partially overlapping materials primarily originate from the experimental structures in the inorganic crystal structure database (ICSD)~\cite{ICSD} and belong to approximately 8,000 observed structural prototypes~\cite{ye2022novel}. Aside from the original structures in the ICSD, several databases contain entries derived from them via the substitution of chemical elements and subsequent geometry optimization~\cite{hautier2011data, Schmidt2023, weiTCSPTemplateBasedCrystal2022, Merchant2023,kirklin_open_2015}.
This represents biased screening, which, by construction, will be limited in its ability to find crystal structures that are vastly different from those already known~\cite{hicks_aflow-xtalfinder_2020}.

In this work, we employ a \emph{computationally tractable and highly efficient} framework that scales structure exploration by orders of magnitude while retaining \emph{unbiased and systematic} coverage of configuration space. This allows us to overcome the limitations of the structural prior inherited from existing databases and makes it feasible to discover new prototypes and low-energy structures in unexplored regions of the massive structural space at the scale of tens of billions of protostructures.

\begin{figure*}[!htbp]
	\centering
	\includegraphics[width=1.0\textwidth]{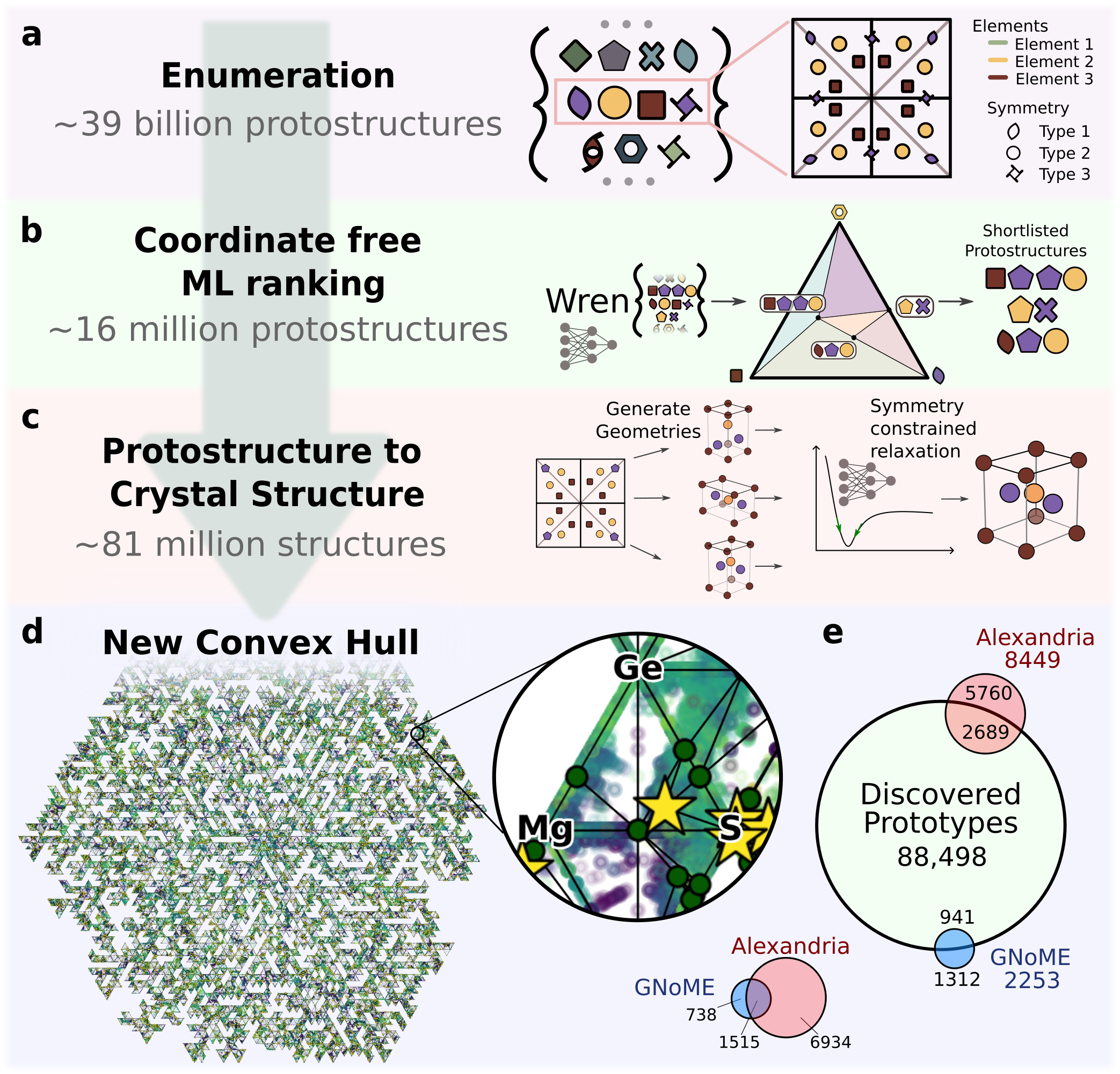} 
	\caption{\textbf{Illustrated workflow}: (a) We systematically enumerate over 39 billion protostructures for binary and ternary compounds with elements from lithium to bromine; (b) Using an updated version of Wren, an ML model trained on the Alexandria database, we construct a prescreen-level convex hull of thermodynamical stability which allows us to identify 0.04\% of the protostructures as relevant for the next screening stage; (c) Each of the remaining protostructures are concretized as multiple crystal structures, and the MACE and ORB MLIPs are used to identify local minima, producing \TotalCrystalStructures\ optimized structures; (d) From these structures a cumulated ``web'' of \NumberOfPhaseDiagrams\ ternary phase diagrams is constructed. Each yellow star in the phase diagrams represents a structure whose protostructure label is not part of the Alexandria dataset, but is predicted to be stable by our workflow; (e) The phase diagrams allow the identification of \ProtoStructuresUnderHundred\ structures with a decomposition enthalpy below 100 meV/atom of the convex hull of thermodynamical stability and spans 88,498 \emph{new prototypes} not in Alexandria or the GNoME data set~\cite{Merchant2023}. The Venn diagrams show the number of prototypes in the different sets within the same structural limits, 5 atoms in the asymmetric unit and 64 atoms in the unit cell, to show the large structural diversity resulting from our screening.}
	\label{fig:intro_figure} 
\end{figure*}

Our framework consists of two steps: first, a coarse-grained prescreening using protostructures, an enumerable representation of crystal structures where every grain itself encompasses a vast structural subspace \cite{goodall_rapid_2022, Parackal2024}; and, second, an in-depth exploration of this structural subspace using recently developed ML methodology. The two-step process, combined with a highly optimized GPU/CPU implementation, enables screening at the extreme scales required to largely overcome the combinatorial bottleneck, as illustrated in Fig.~\ref{fig:intro_figure}. In an already well-explored space of binary and ternary crystal structures with elements in the range lithium to bromine, we identify \ProtoStructuresUnderHundred\ crystal structures in the energy range close to thermodynamical stability usually considered as relevant for (meta)stability ($100\ \mathrm{meV/atom}$) spanning \PrototypesUnderHundred\ (of which, novel \PrototypesUnderHundredNovel\/) crystal prototypes.

In more detail, the configurational space of a crystal structure with $N$ atoms in the unit cell 
comprises $6+3N$ continuous degrees of freedom from the lattice and coordinates.
Prior work discussing ways to discretize and enumerate the structural space goes back at least to Frank and Kasper, who in 1958 discussed the description of crystal structures as triangulated polyhedra~\cite{Frank1958}. Since then, works have explored various restricted enumerations of graphs representing crystal prototypes called crystal nets (see Ref.~\cite{DelgadoFriedrichs2003} for an overview). More recent developments, including our own works, discuss various ways to enumerate the full space~\cite{gusevOptimalityGuaranteesCrystal2023a, goodall_rapid_2022,Parackal2024}. However, so far, there has been no convincing practical demonstration of how to address the resulting combinatorial explosion in general systematic screening. 

In our approach, the degrees of freedom of the crystal structures are reduced by using a coarse-grained representation, protostructures, that only specify the type of symmetries for the atoms in the structure. 
The second stage of screening uses a fine-grained representation that adds the degrees of freedom of the lattice and each of the atomic positions~\cite{Wang2018, Reinaudi2000, lenz2019parametrically}.
The resulting combinatorial enumeration 
scales far better to complex structures than the original configurational space.
Some prior works have applied a similar approach to the CSP problem, but then addressed the remaining combinatorial challenge by the use of statistics from existing crystal structure databases to prioritize the most common existing combination of chemical elements and structural configurations \cite{hybrid, ward_three_2015,changShotgunCrystalStructure2024, wangSLICESPLUSCrystalRepresentation2024}. While this has generally been a successful strategy, it ultimately leads back to the kind of bias from existing data that we seek to avoid.

There is also an increasing interest in applying generative AI to overcome the combinatorial barrier in the crystal structure prediction problem. Diffusion, transformer, or similar models are trained on existing data and then used to generate completely new crystal structures, see Ref.~\cite{DeBreuck2025} for an overview. However, since these models generate structures out of a generalized distribution spanned by the training data, the approach is, by design, biased towards that training data, i.e., the known parts of the structural space. The extended set of materials resulting from our screening can be used to expand the reach of such methods.

\subsection*{Coarse-grained representation}

Central to this work is the coarse-grained \emph{protostructure representation} of crystal structures \cite{goodall_rapid_2022, Parackal2024}. All crystal structures can be classified into 230 space groups. Sets of points in the unit cell that are equivalent under the symmetry operations of a given space group define its Wyckoff positions \cite{ITA2002}. These positions are labeled using letters (a, b, ...) that, in reverse order, reflect how strongly a site is constrained by symmetry and dictate the number of symmetry-imposed copies of that position in the unit cell. A protostructure specifies a crystal structure only in terms of its space group, its chemical elements, and the letters of the Wyckoff positions they occupy. 

\begin{figure}
	\centering
	\includegraphics[width=0.9\linewidth]{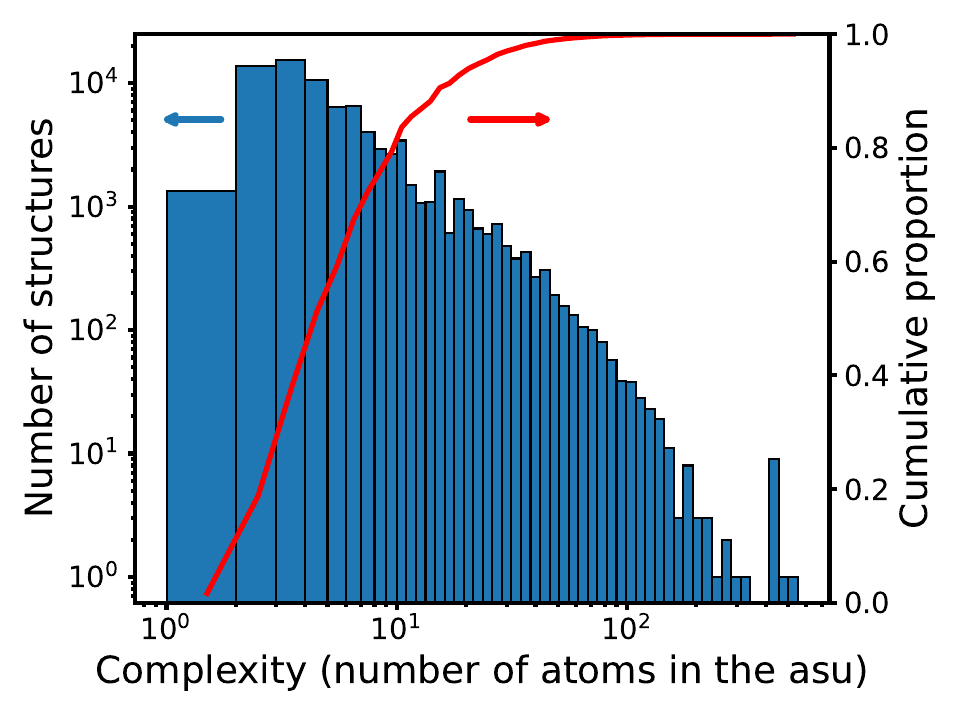}
	\caption{Distribution of ordered crystal structures with space group $> 2$ in the ICSD over structural complexity (the number of atoms in the asymmetric unit). A log-log plot is used to show the long tail nature of the distribution. If these structures, sourced from publications, are representative of the space of discoverable crystal structures, it appears \emph{complexities up to just five cover more than half of the complete relevant structural space}, i.e., the level exhaustively explored in this work. Extending our approach within reasonable levels of computational expense should allow reaching at least $80-90\%$.\label{fig:complexity}}
\end{figure}

Each occupied Wyckoff position corresponds to a non-equivalent atom of the structure. Those atoms have precisely one equivalent copy that belongs to the asymmetric unit (the smallest simply connected part of space that via the symmetry operations fills the whole space). We will refer to the number of non-equivalent atoms as the \emph{complexity} of a structure since more atoms in the asymmetric unit correlate with larger, more structurally diverse unit cells. As illustrated in Fig.~{\ref{fig:complexity}}, more than half of the complete space of the primarily experimentally observed crystal structures collected by the ICSD~\cite{ICSD} are at, or below, complexity five. The increasingly long-range order that structures of larger complexity need to adhere to will also be less likely to hold at higher temperatures. Hence, the entropy contribution imposes limits on how high the complexity can go for ordered structures at room temperature.

\subsection*{Stability}

The concept of stability is complex even in the domain of ordered single crystals with moderately sized symmetrical unit cells, which is the most common context of the CSP problem. It spans at least thermodynamic, dynamical, and mechanical criteria. Nevertheless, to map out composition phase diagrams, we will largely follow the convention of primarily considering the decomposition enthalpy $H_d$ relative to a convex hull spanned by the formation energies of a set of known competing phases: $H_d$ is the difference of the formation energy of the material, $H_f$, and that of the most favorable linear combination of competing phases at the same composition $H_{f,\mathrm{min}}$. The lower $H_d$ is, it is regarded as more likely for the material to be stable or metastable and possible to synthesize \cite{Bartel2019}. Since $100\ \mathrm{meV/atom}$ is the approximate formation energy error of density functional theory (DFT) calculations for general chemistries and structures at the most common level of theory~\cite{kirklinOpenQuantumMaterials2015}, a cutoff of $50 - 100\ \mathrm{meV/atom}$ is often used to designate materials relevant for further investigation~\cite{Bartel2019}. 
\subsection*{Overview of screening workflow}

The Wyckoff Representation Network (Wren) is a message-passing neural network that, given a protostructure, predicts the formation energy of the corresponding crystal structure with optimized atomic positions, i.e., the relaxed structure, at an accuracy comparable with models that utilize the full (already relaxed) $6+3N$ representation~
\cite{goodall_rapid_2022, riebesellMatbenchDiscoveryFramework2024}. We have shown that this model is predictive 
even for protostructures that belong to prototypes that are not part of the training data \cite{Parackal2024}. Hence, Wren is perfectly adapted for a prescreening crawl through the combinatorial space of protostructures to assess structural stability based on relative formation energies to extract the small subset relevant for the second stage of screening.

We have reimplemented Wren as a transformer model\footnote{The change of architecture was inspired by work by J.~Riebesell~\cite{riebesellMatbenchDiscoveryFramework2024}} using the JAX Python library for hardware accelerators~\cite{jax2018github}, reducing the computational cost by up to a factor of 40  of the original GPU implementation~\cite{goodall_rapid_2022}. It is now capable of evaluating \emph{millions of protostructures per second on a current-generation GPU} (see Methods for more details). We combine this qualitative leap in performance with restrictions of the chemical space to the range of lithium to bromide (excluding the Noble gases), a cut-off in the structural space of only binary and ternary compounds with Wyckoff complexity $\leq 5$, and the fixed limit of 64 atoms in the unit cell. With fairly limited computational resources (detailed below), we are able to exhaustively pre-screen all the \emph{39 billion protostructures} within these restrictions, of which 15 million are retained based on the energy ranking provided by Wren,  as illustrated in Fig.~\ref{fig:intro_figure}a and b.

In the second stage of screening, each identified protostructure is concretized into crystal structures many times over by assigning values to the degrees of freedom allowed by the Wyckoff positions using Latin hypercube sampling \cite{loh1996latin}  (for details, including how the number of samples is determined, see Methods). The resulting structures are optimized under constrained symmetry using a machine-learning interatomic potential (MLIP), our primary choice is a recent version of MACE~\cite{mace,Batatia2025ACEsuit} (MPA-0) but with a single validation of the final energy using ORB \cite{orb} (see Methods).
Filtering and concretization turn the 39 billion protostructures into \TotalCrystalStructures crystal structures from which \NumberOfPhaseDiagrams\ ternary phase diagrams can be created, as illustrated in Fig.~\ref{fig:intro_figure}c and d.

\begin{figure*}[!htbp]
	\centering
	\includegraphics[width=0.9\textwidth]{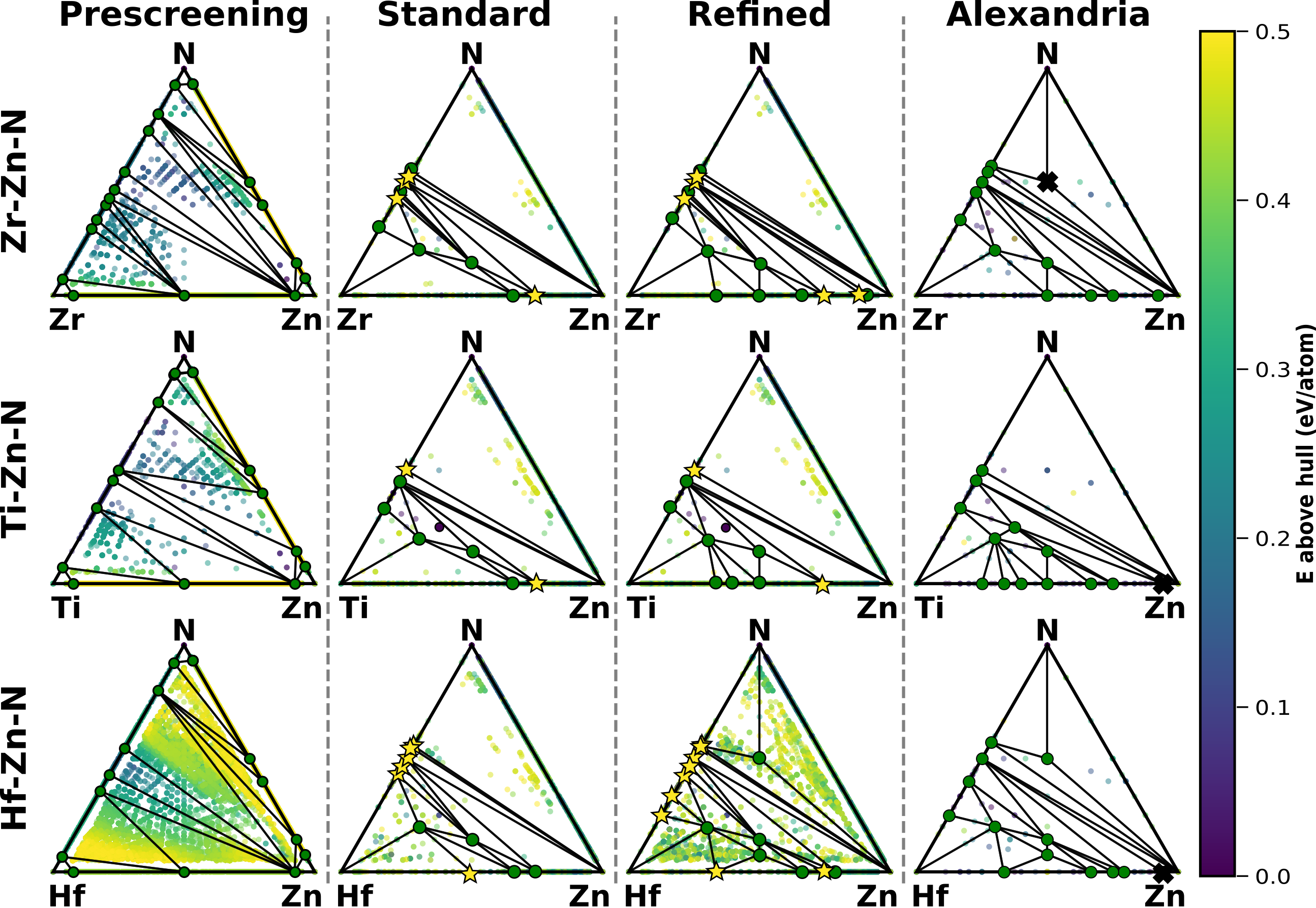} 
	\caption{\textbf{Validation} for the materials systems Hf-Zn-N, Ti-Zn-N, Zr-Zn-N with columns: \emph{Prescreening} for phase diagrams from enumeration and prescreening-level ML formation energy predictions; \emph{Standard} for the standard level of our second stage of concretization and MLIP-relaxation; \emph{Refined} for extended concretization beyond 5000 structures per phase diagram; \emph{Alexandria} for the well-explored diagrams from structures known before this work. Phases in Alexandria not found by our workflow are shown as black x and a phase in Alexandria offset by the changes in thermodynamical stability due to our new phases is shown as a black dot.  New phases not in Alexandria are shown as yellow stars.\label{fig:taata_0}
		}
\end{figure*}

\subsection*{Demonstration on known systems}

The ternary and binary systems in the material systems Hf-Zn-N, Ti-Zn-N, Zr-Zn-N have been thoroughly investigated by prior high-throughput DFT calculations, which makes them a robust reference for the validation of our methods \cite{tholander2016strong}, which will be done in two ways. 
First, we use the same methodology as for the screening of the larger range of chemistries below. Second, we extend the investigation in a targeted effort possible for a focused screening in a more narrow chemical space or if more computational resources are available for a broader screening. 

For each chemical system, we enumerate all symmetry-distinct protostructures within the complexity limit of five, predict their formation energies using the Wren model, and retain those that are stable or close to stable (see Methods).
The phase diagrams from this `Prescreen' stage are shown in Fig.~\ref{fig:taata_0}. The resulting structures are concretized and MLIP-relaxed using the methodology described in the Methods section.
The adjusted phase diagrams are shown in the `Screening' column of Fig.~\ref{fig:taata_0}.
Finally, we release the specific limitation in the screening of retaining only the 5000 structures with the lowest decomposition enthalpy $H_d$ per phase diagram and keep all protostructures with a $H_d$ relative to the Alexandria database of less than $100\ \mathrm{meV/atom}$. This results in 66,180; 37,926; and 60,600 protostructures respectively for the Hf-Zn-N; Ti-Zn-N; and Zr-Zn-N systems. These structures are concretized and MLIP-relaxed to give the `Refined' column of Fig.~\ref{fig:taata_0}. When comparing the resulting phase diagrams, we stress that these chemical systems were chosen for validation because they are particularly well explored. Hence, the objective is to rediscover known phases rather than to find new ones, suggesting that the methods are predictive to the same degree in less-explored systems. Furthermore, in this comparison, two phases are considered to be the same or different based on their protostructures. 

It is clear that the second `Screening' stage is crucial for achieving the precision in formation energy necessary for these diagrams. There is less difference between the `Screening' and `Refined' steps, with only a few new binary phases and one ternary phase added. The binary phases are in ranges of multiple structures with similar energies, so their inclusion or omission does not significantly influence the diagrams. However, the single ternary phase, $\mathrm{Hf}_{6}\mathrm{Zn}_{6}\mathrm{N}$, is important and was omitted due to the limit of 5000 structures, motivating the expense of the extra computational effort when possible. 

A few phases in `Alexandria' are missing even in `Refined' (shown as black x markers). All except two of them are outside the screened complexity range, these are (with complexities): $\mathrm{Hf}_{3}\mathrm{N}_{4}$ (7), $\mathrm{Hf}\mathrm{Zn}_{16}$ (8), $\mathrm{Ti}\mathrm{Zn}_{16}$ (8), $\mathrm{Zr}_{3}\mathrm{N}_{4}$ (7).
One of the omitted phases within the screened complexity range of five is $\mathrm{Zr}\mathrm{Zn}\mathrm{N}_{2}$. Its energy is estimated by Wren in the prescreening to have a decomposition enthalpy of $164 ~\mathrm{meV/atom}$ and therefore is not taken to the second stage. The second omitted phase, $\mathrm{Ti}_2\mathrm{Zn}\mathrm{N}$ is in our phase diagram but has ended up just slightly ($13\ \mathrm{meV/atom}$) above our convex hull, and thus was not included on our hull because it is displaced, but only barely, by the nearby phases.
Conversely, our methods uncover 19 binary protostructures (6 at new compositions) missing from the `Alexandria' diagram, but since they are in binary regions of several competing phases none of them are crucial for the general features of the diagrams.

\begin{figure*}[!htbp]
    {\centering
        \textbf{\sffamily a}~\includegraphics[height=35 ex,valign=t]{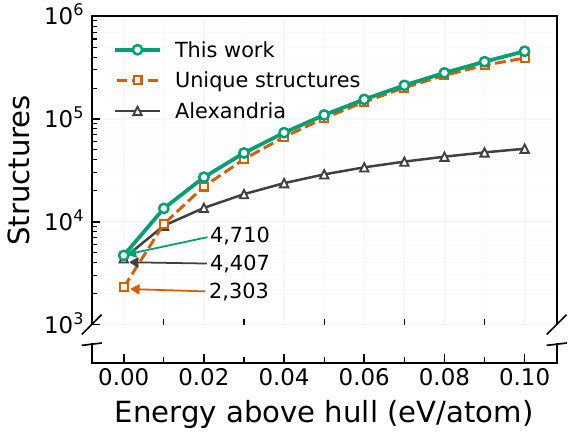}~~
        \textbf{\sffamily b}~~\includegraphics[height=33 ex,valign=t]{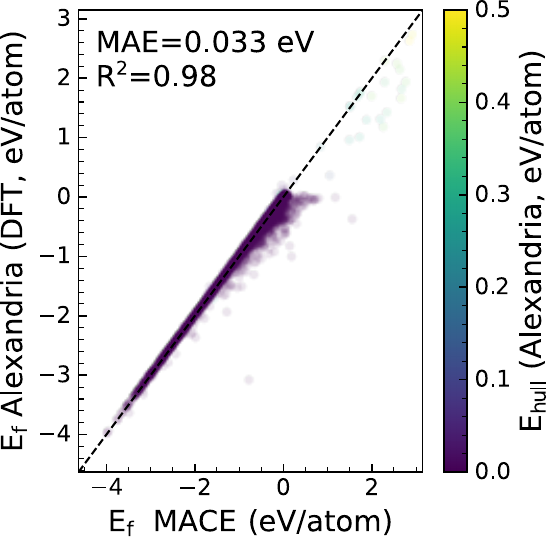}
    
    }

	\caption{\textbf{Large-scale screening outcome and error analysis:} (a) Comparison of the number of unique structures discovered in our screening (red squares) out of all we identified (green circles) vs.\ the number of structures known in Alexandria within the screened chemical space, but with no other limits on complexity or number of atoms (gray triangles). The graphs show the cumulative count of structures with a decomposition enthalpy $H_d$ \emph{below} the value shown on the x-axis. It is common to consider structures with a $H_d$ below $50-100\ \mathrm{meV/atom}$ as relevant, to accommodate the accuracy of the underlying DFT computations and the possibility for metastability \cite{Bartel2019}.Unique structures are structures whose protostructures are not in Alexandria, and for which $H_d \leq 100\ \mathrm{meV/atom}$ relative to the convex hull of stability of the union of our structures and Alexandria. For the full range up to $100\ \mathrm{meV/atom}$, we discovered \ProtoStructuresUnderHundred\ materials spanning \PrototypesUnderHundred\ new prototypes; (b) An error analysis for the predicted materials also present in Alexandria. 
}
    \label{fig:after_mace_stastitics}

\end{figure*}

\subsection*{Massive-scale screening}
\label{sec:screening}
For the full screening illustrated in Fig.~\ref{fig:intro_figure}, we systematically enumerate all symmetry-distinct protostructures for binary and ternary compounds with elements in the range lithium to bromine for 
up to a complexity of five and with less than 64 atoms in the unit cell, giving 
1,031,440,680 binary protostructures, and 38,326,141,030 ternary protostructures across all 4960 (4495 ternary and 465 binary) phase diagrams. 
Applying the Wren prescreening workflow (see Methods) results in 15.7 million protostructures at a computational expense of roughly 16,000 GPU hours on a single supercomputer node with eight AMD MI250x GPUs.
The second stage of the screening concretizes and optimizes each protostructure over its respective degrees of freedom, invoking over a billion local MLIP minimizations at the computational expense of ~170,000 GPU-hours, using the same GPUs.

The result is 71,692 structures, which, based on their decomposition enthalpies (calculated using the MACE MPA-0 MLIP\cite{Batatia2025ACEsuit}), are at or below the Alexandria convex hull, of which \StableProtostructuresThatDoNotExitInAlexandria\ belong to protostructures not already in Alexandria. However, as discussed, it is common to consider structures with decomposition enthalpies up to $50-100\ \mathrm{meV/atom}$ as candidates for synthesis due to the combination of the accuracy of the computational methods and the possibility of metastability \cite{Bartel2019}. As a function of increasing positive decomposition enthalpies, the number of new relevant structures we have discovered grows to \ProtoStructuresUnderHundred\ for $H_d \leq 100\ \mathrm{meV/atom}$. These results are shown in Fig.~\ref{fig:after_mace_stastitics}a (with data in Supplementary note 7). Furthermore, a more in-depth analysis reveal that these structures span \PrototypesUnderHundred\ unique crystal prototypes. This number highlights the structural diversity of the structures discovered in this work: within the same structural limits (max complexity of five, and max 64 atoms in the unit cell) Alexandria has 8449 unique prototypes and GNoME 2253. A more in-depth comparison is provided in Supplementary note 5.

So far, energies have been evaluated with MLIPs. As a further validation of the results, we executed high-throughput DFT on the \StableProtostructuresThatDoNotExitInAlexandria\ new structures with $H_d \leq 0$  and confirmed that 631 of them remain below zero (i.e., on the convex hull of stability). The other structures with positive $H_d$ remain within $25\ \mathrm{meV/atom}$ of the hull, which in relation to the accuracy of DFT and the precision of other methods involved, are interpreted as passing validation. 
A broader statistical analysis of the errors is shown in Fig.~\ref{fig:after_mace_stastitics}b, where the mean absolute error of the final MLIP energy predictions relative to the DFT calculations in Alexandria is 33 meV/atom, which includes both the error of the MLIP in reproducing DFT energies and the error caused by the concretization workflow not finding the absolute minimum energy configuration.
There is a bias in the errors toward higher formation energies, suggesting that our methodology is more likely to yield false negatives than positives.

\subsection*{Conclusion}
This work has presented a systematic and practical workflow for the massive enumeration and screening of the space of crystal structures. By utilizing enumeration limited by structural complexity in terms of the number of atoms in the asymmetric unit, and leveraging the Wren ML model for rapid prescreening, we could efficiently evaluate billions of unique protostructures. Our approach has been validated both using the well-explored Hf-Zn-N, Ti-Zn-N, and Zr-Zn-N material systems and by comparing the outcome of the broader screening against known structures in the Alexandria database, demonstrating high recovery rates within the imposed complexity limits.

The outcome of this work is \TotalCrystalStructures\ crystal structures force converged using the MACE MLIP, 
from which we find half a million in the range of negative or low decomposition enthalpies.
Of particular importance is that they span 88 thousand new prototypes, representing a true qualitative shift in structural diversity for materials databases.
These results highlight the effectiveness of symmetry-guided enumeration combined with ML models for large-scale materials discovery.

Finally, despite the massive extent of the space screened so far (39 billion protostructures) our methodology provides a scalable foundation for future studies. There is a major asymmetry in computational expense, with more than 90\,\% spent on the concretization and MLIP relaxation stages. Hence, leveraging the progress made in this work to retrain our prescreening ML model for increased precision will allow fewer systems to propagate to the expensive second stage, enabling the exhaustive screening of the space of crystal structures to reach higher structural complexity for larger parts of the periodic table.

\section*{Acknowledgments}

\paragraph*{Funding:}
We acknowledge financial support from the Swedish Research Council (VR) through Grant No. 2020-05402 and the Swedish e-Science Research Centre (SeRC).
F.T. acknowledges support through ERC
Grant (UNMASCC-HP, 101117758) and the Swedish government’s Strategic Research Area in Materials Science on
Functional Materials at Linköping University (faculty grant SFO-Mat-LiU 2009-00971).

The computations were enabled by resources provided by the National Academic Infrastructure for Supercomputing in Sweden (NAISS), partially funded by the Swedish Research Council through grant agreement no. 2022-06725. We also acknowledge NAISS for providing access to
the LUMI supercomputer, owned by the EuroHPC Joint
Undertaking and hosted by CSC (Finland) and the LUMI
consortium.
\paragraph*{Author contributions:}
\textbf{Abhijith S Parackal:} Methodology, Investigation, Formal analysis, Data curation, Visualization, Validation, Software, Writing – original draft, Writing – review \& editing; \textbf{Florian Trybel:} Software, Resources, Methodology, Writing – review \& editing; \textbf{Felix Faber:} Conceptualization, Supervision, Methodology, Writing – review \& editing; \textbf{Rickard Armiento:} Conceptualization, Supervision, Software, Resources, Methodology, Funding acquisition, Writing - review \& editing. 

\paragraph*{Data and materials availability:} at the time of publication, the discovered materials are available via our open access database at \url{https://anyterial.se}, and the source code for \emph{httk} and the \emph{httk-symgen} module are available via our GitHub organization page \url{https://github.com/httk}.

\bibliography{main.bib}
\bibliographystyle{unsrt}

\clearpage

\section{Methods}

\subsection*{Improved Wren Implementation}

The original Wren implementation available through the Aviary software package \cite{aviary,goodall_rapid_2022} uses a message-passing neural network to predict the lowest formation energy configuration allowed by the degrees of freedom for the occupied Wyckoff positions.
An ensemble of 10 models, each initialized differently, provides both improved accuracy and an internal measure of uncertainty that combines aleatoric and epistemic components.
In this work, we use a completely reimplemented variant of the Wren model based on the transformer architecture \cite{riebesellMatbenchDiscoveryFramework2024} using JAX \cite{jax2018github} and substantial modifications to the data loading and training workflow, achieving up to a 40-fold increase in inference throughput.
Furthermore, we added point group information for Wyckoff positions as an additional descriptor to the original Wren model. The compactness of the point group (32 point groups vs.\ 1732 Wyckoff positions) enables the model to perform better in regions outside the training data. 
The model is trained on a version of the ``PBE 3D dataset'' of the Alexandria materials database dated 2023.12.29, which contains 4,5 million structures~\cite{Schmidt2023}.
Our implementation relies heavily on symmetry detection and other symmetry-related functionalities that uses spglib \cite{spglib} and data from the Bilbao Crystallographic Server \cite{bilbao}.

\subsection*{Volume Estimation}

To accurately estimate initial unit cell volumes used in the concretization stage of the workflow, we have developed a linear regression model that predicts effective atomic radii based on chemical identity and coordination environment. This approach addresses the limitations of fixed atomic radii, as the effective size of atoms varies significantly with their local chemical environment and bonding characteristics. Supplementary note 1 Fig.~\ref{fig:radii_comparison} gives details on the comparison between our fitted effective atomic radii and standard literature values (covalent and van der Waals radii), demonstrating the improved accuracy of our volume estimation approach for crystal structure prediction.

\subsection*{Workflow details}

As outlined in the previous sections, our screening comprises two main stages: (i) enumeration of Wyckoff protostructures and prescreening with Wren to identify candidates near thermodynamic stability, and (ii) symmetry-aware realization and relaxation of explicit crystal structures using MLIPs, followed by validation and storage of the resulting low-energy structures for downstream analysis. 

\paragraph{Enumeration}

Our enumeration covers all binary and ternary protostructures up to complexity five, containing all elements of the periodic table between lithium and bromine, resulting in $\sim 39$ billion protostructures. Wren \cite{goodall_rapid_2022}, trained on the Alexandria dataset ($\sim$ 4M structures; see Supplementary note 3 for more details), is then used to prescreen these protostructures rapidly.

\paragraph{Prescreening}
This step involves stability calculation of the enumerated protostructures using the Wren-predicted formation energies.
A convex hull of stability is constructed from the Wren energies using the ASE Python library \cite{ase-paper}.
Subsequently, we construct a reference convex hull from the Alexandria dataset using uncorrected DFT formation energies, thereby applying the same energy definition to both datasets (i.e., without any post-hoc corrections, see \cite{ongPythonMaterialsGenomics2013a, mace}).
As Wren is trained on the Alexandria dataset, its predicted formation energies are on the same reference scale as the DFT formation energies and can therefore be compared directly for stability assessment.

The Alexandria convex hull serves as an upper bound for screening, i.e, protostructures that Wren predicts to lie more than a small threshold (100 meV/atom) above the Alexandria convex hull are discarded. Consequently, the region of interest consists of protostructures predicted by Wren to fall below the Alexandria convex hull, with the lower bound defined by the Wren-predicted convex hull itself.

For computational feasibility, we forward at most 5000 protostructures per chemical system to the subsequent structure-realization stage, resulting in a total of approximately 15 million protostructures.

\paragraph{Missing Structures and Recovery Efficiency} For the binary and ternary chemical systems comprising elements from lithium (Li) through bromine (Br), excluding the noble gases, a total of 4,960 unique element combinations exist.
Following Wren screening, candidate structures are ranked by the predicted energy above the convex hull given by Wren.
Analysis of the Alexandria dataset reveals that 2,267 out of the 4,960 possible systems (45.7\%) host at least one thermodynamically stable phase, totaling to 4,999 stable structures.
For comparison against the workflow, we applied identical complexity and atomic‐count restrictions to the Alexandria dataset and recalculated the stability of this collection.
 Under these constraints, 1,797 systems (36.2\%) contain at least one stable phase, totaling 3,200 stable phases. 

We then evaluated the coverage of our screening protocol by determining the number of candidates that must be examined when ordered by energy above the predicted hull
 to recover the full set of restricted‐Alexandria stable phases.
With a per‐system computational budget of 5,000 candidate structures, the Wren screening retrieves 2,676 of the 3,200 restricted‐Alexandria stable phases,
 corresponding to an overall recovery rate of 83.6\%, as illustrated in Supplementery note 2, Fig.~\ref{fig:cumulative_recovery}.

\paragraph{Latin hypercube sampling over symmetry-allowed degrees of freedom} After constructing the Wren convex hull, we obtain a prioritized list of protostructures predicted to be thermodynamically favorable.
The next step is to instantiate explicit crystal structures from these abstract protostructure labels.
This process is non-trivial, as each protostructure encodes only the symmetry (space group and Wyckoff positions) and composition, but not the explicit atomic coordinates or lattice parameters.

To generate candidate structures, we follow a workflow similar to that described in \cite{Parackal2024}, with notable changes in the instantiation of lattice parameters.
In short, for each protostructure, we first identify all the degrees of freedom (DOFs), given by the sum of the Wyckoff and lattice degrees of freedom.
To efficiently sample the high-dimensional configurational space, we employ Latin Hypercube Sampling \cite{loh1996latin, mckay2000comparison, 2020SciPy-NMeth} (LHS), which ensures a more uniform and representative coverage compared to random sampling.
The number of initial structures generated per protostructure is determined as a heuristic function of the number degrees of freedom $d$, rescaled by the number of atoms $N$, given by
\[
N_{\mathrm{structures}} = \min\!\left(20 + \frac{10\, d^{1.3}}{N^{0.5}},\, \frac{3000}{N}\right)
\]
The upper bound of 3000 enforces a fixed atoms budget adapted for the hardware we use, and prevents out-of-memory errors.

The LHC sampling requires defining the bounds of the space.
For Wyckoff positions, each coordinate satisfies \(x_i \in (0,1]\).
Lattice degrees of freedom are unbounded and require careful consideration in defining the bounds to avoid unphysical, highly strained, or sparse structures.
To address this, we estimate the expected unit-cell volume for each composition using a simple linear regression model trained on the Materials Project database to predict effective atomic radii for a given atom (see Volume Estimation above).
The initial lattice parameters are then sampled such that the total cell volume varies randomly within $\pm$10\% of the predicted value.
For non-cubic systems, we further constrain the lattice vector ratios (e.g., $a/b$ and $b/c$) within physically reasonable bounds (in the present work, we set a maximum $ a/b$ ratio of 14), ensuring that both elongated and compressed cells are explored. 

A key requirement of our workflow is that all relaxations must preserve the symmetry of the original protostructure.
While some structures may relax into higher-symmetry (supergroup) configurations, they must remain mappable to the original Wyckoff assignment to maintain consistency in the enumeration.

The structure realisation and relaxation pipeline proceeds in three steps:

\paragraph{Step 1: Distance Optimisation:} Where structures initialised with LHC sampling are first optimised to ensure that the interatomic distances are physically reasonable.
This is achieved by minimising the distance penalty function defined as:
\begin{equation}
	C_{d} = \frac{1}{2}\sum_{i=1}^{n}\sum_{j=1}^{n}\max\left(0,\;D^{\mathrm{min}}_{(i,j)} - \left(D_{(i,j)} + d_{\mathrm{lat}}^{\mathrm{min}}\delta_{ij}\right)\right)
\label{eqn:distance_eq}
\end{equation}
where $D_{(i,j)}$ is the shortest periodic distance between atoms $i$ and $j$, and $d_{\text{lat}}^{\min}$ is the shortest nonzero lattice translation included to avoid self-interaction artifacts.
While, in principle, a robust interatomic potential should be able to relax structures even when atoms are initialized close together, in practice, we find that starting from physically reasonable atomic separations leads to more reliable and efficient relaxations \cite{landscapepaper}.
Therefore, we ensure that all initial structures have sensible interatomic distances before proceeding to the next stage.

\paragraph{Step 2: Symmetry-Constrained Geometry Optimisation:}  
The distance-optimised structures are then relaxed using the \macempa\ MLIP \cite{mace}. This potential is trained on (uncorrected) selected set from Materials Project (MPtraj)\cite{jainCommentaryMaterialsProject2013, chgnet} and Alexandria DFT energies, which means that the energies across Wren, MACE, and the DFT values in Alexandria, are directly compsarable.
We adopt the symmetry-constrained relaxation formalism of Lenz et al. \cite{lenz2019parametrically}, where atomic forces are projected onto the subspace of symmetry-allowed degrees of freedom.
 Unlike the original implementations, we leverage JAX's automatic differentiation to efficiently compute the Jacobian for an arbitrary lattice that includes an angular degree of freedom.
The relaxation is performed using the Adam method \cite{kingmaAdamMethodStochastic2017,deepmind2020jax}, allowing for fast convergence while strictly preserving the original space-group symmetry.

\paragraph{Step 3: Final Validation and Filtering.}  
The lowest-energy relaxed structures from Stage 2 are further refined using the ASE library \cite{ase-paper}, again employing the \macempa.
We utilize the functionality in the ASE Python library \cite{ase-paper} to modify a cell and atomic positions using a filter implemented in \texttt{FrechetCellFilter} and \texttt{FixSymmetry} to guarantee that symmetry is preserved throughout the final relaxation. We use a force tolerance of \(10^{-3}\,\mathrm{eV/\AA}\) for convergence. 
For additional validation, we perform a static energy calculation using the \orb\ model \cite{orb} as an independent verification.
Only structures for which the \macempa\ and \orb\ models predicted energies agree within $100\ \mathrm{meV/atom}$ are retained, providing a measure of confidence in the reliability of the relaxation.
Further, we ensure that the minimum interatomic distance is maintained after the final relaxation as well.
We retain the five lowest energy crystal structures for a given protostructure in our database, as well as those structures that have shown force-convergence yet failed to pass the distance check for future reference. However, when counting discovered structures, we only count each protostructure once, i.e., its lowest energy minimum.
As a final step, we rerun the symmetry finder (spglib \cite{spglib}) to detect cases where the process has, e.g., increased symmetry, to ensure to correctly count structures unique in this work (the red line in Fig.~\ref{fig:after_mace_stastitics}). As a result, a small number of final structures (94) do not respect the initial limit of 64 atoms per unit cell. All validated structures are stored in a duckdb \cite{duckdb} file for subsequent analysis.

\clearpage
\onecolumn
\setcounter{figure}{0}
\setcounter{table}{0}
\makeatletter 
\renewcommand{\thefigure}{S\@arabic\c@figure}
\renewcommand{\thetable}{S\@arabic\c@table}
\makeatother
\section*{Supplementary Information for: Screening of 39 billion protostructures for materials discovery}

Abhijith S Parackal, Florian Trybel, Felix Andreas Faber, Rickard Armiento

\subsection*{Supplementary note 1: Protostructure label format}
As explained in the main paper, a \emph{protostructure} is a generalized representation that assigns specific chemical elements to the occupied Wyckoff sites of a structure while leaving the underlying atomic coordinates unspecified \cite{Parackal2024}.  A \emph{protostructure label} is a convenient format for expressing protostructures. It follows the AFLOW \cite{AFLOWLibraryCrystallographic} labeling scheme, which consists of the anonymous formula, the Pearson symbol, the space group number, and the Wyckoff letters, and it becomes a standardized identifier once a colon (``:'') with an ordered list of elements seperated with dashes (``-'') is appended to indicate which species occupy which Wyckoff position.
For example, NaCl in space group 225 corresponds to the protostructure label 
\texttt{AB\_cF8\_225\_a\_b:Na-Cl}, which uniquely encodes both the structural template and its elemental decoration. 

In the implementation of our full workflow, we utilize the implementation in the Aviary software package \cite{aviary} to convert between structures, protostructures, and prototypes. We also follow the convention in this software for the normalization of protostructures by selecting among all symmetry-equivalent protostructure labels generated through affine-normalizer transformations the one that appears first when sorted by the summed alphabetical indices of its Wyckoff letters and then lexicographically, ensuring an origin-independent and canonically defined protostructure.

\subsection*{Supplementary note 2: Detailed results from the demonstration on Hf-Zn-N, Ti-Zn-N, and Zr-Zn-N}

Table~\ref{tab:hztnzn_table} shows a list of stable protostructures identified in our workflow for the Hf–Zn–N, Ti–Zn–N, and Zr–Zn–N materials systems.
For each stable structure, we report the energy above the convex hull for the corresponding protostructure in the Alexandria database if it is present, as well as the energy above hull of the lowest-energy structure at the same composition in Alexandria.
Entries marked as `Unseen' indicate cases where the corresponding protostructure or composition is not present in the Alexandria dataset.
\begin{table}
	\centering
	\small
	\caption{Stable protostructures from our workflow and the energy above hull for the same protostructure in Alexandria (when available) and for the lowest-energy structure at the corresponding composition.}
	\begin{tabular}{|l|r|r|r|}
    \hline
    \makecell{\textbf{Protostructure}\\\textbf{label}} &
    \makecell{\textbf{Composition}} &
    \makecell{$\boldsymbol{E}_{\mathrm{above\,hull}}$\\\textbf{Lowest protostructure}\\\textbf{in Alexandria}} &
    \makecell{$\boldsymbol{E}_{\mathrm{above\,hull}}$\\\textbf{Lowest composition}\\\textbf{in Alexandria}} \\
    \hline
	\multicolumn{4}{|c|}{\textbf{Hf-Zn-N}} \\
    \hline
     A6BC6\_cF104\_227\_f\_a\_de:Hf-N-Zn & Hf6NZn6       & 0.0                               & 0.0                            \\
     A6B23\_cF116\_225\_e\_ad2f:Hf-Zn   & Hf6Zn23       & 0.0                               & 0.0                            \\
     A3BC3\_cF112\_227\_f\_c\_de:Hf-N-Zn & Hf3NZn3       & 0.0                               & 0.0                            \\
     A3BC\_oP20\_62\_cd\_a\_c:Hf-N-Zn    & Hf3NZn        & Unseen                            & 0.0                            \\
     AB2C\_hP4\_156\_b\_ac\_a:Hf-N-Zn    & HfN2Zn        & 0.0                               & 0.0                            \\
     A5B6\_mC22\_12\_agh\_ij:Hf-N       & Hf5N6         & Unseen                            & Unseen                         \\
     A8B7\_hR45\_166\_ch\_ade:Hf-N      & Hf8N7         & Unseen                            & Unseen                         \\
     A4B5\_tI18\_79\_c\_ac:Hf-N         & Hf4N5         & Unseen                            & 0.00824                        \\
     A4B3\_hR42\_166\_abde\_h:Hf-N      & Hf4N3         & Unseen                            & 0.03688                        \\
     AB3\_tP4\_123\_a\_ce:Hf-Zn         & HfZn3         & Unseen                            & 0.0                            \\
     A2B\_tP6\_123\_gh\_ad:Hf-Zn        & Hf2Zn         & Unseen                            & 0.0                            \\
     AB2\_hP6\_194\_b\_f:Hf-Zn          & HfZn2         & 0.00325                           & 0.0                            \\
     A2B\_hP18\_162\_2k\_abh:Hf-N       & Hf2N          & Unseen                            & 0.00346                        \\
     A3B\_hP32\_163\_2i\_bcf:Hf-N       & Hf3N          & Unseen                            & 0.0                            \\
     AB\_hP12\_194\_bf\_af:Hf-N         & HfN           & Unseen                            & 0.0                            \\
    \hline
	\multicolumn{4}{|c|}{\textbf{Ti-Zn-N}} \\
    \hline
     AB3C3\_cF112\_227\_c\_f\_de:N-Ti-Zn & NTi3Zn3       & 0.0                               &                         0      \\
     AB3C\_oC20\_63\_a\_cf\_c:N-Ti-Zn    & NTi3Zn        & 0.0                               &                         0      \\
     A3B2\_tI10\_139\_ae\_e:Ti-Zn       & Ti3Zn2        & 0.0                               &                         0      \\
     A4B5\_tI18\_87\_h\_ah:N-Ti         & N4Ti5         & 0.3455                            &                         0.3455 \\
     AB3\_oP4\_47\_a\_dfg:Ti-Zn         & TiZn3         & Unseen                            &                         0      \\
     A5B6\_mC22\_12\_agh\_ij:N-Ti       & N5Ti6         & 0.0                               &                         0      \\
     A2B\_tI6\_139\_e\_a:Ti-Zn          & Ti2Zn         & 0.0                               &                         0      \\
     AB\_tP4\_123\_h\_ab:Ti-Zn          & TiZn          & Unseen                            &                         0      \\
     AB2\_tP6\_136\_a\_f:N-Ti           & NTi2          & 0.0                               &                         0      \\
     AB\_cP8\_221\_bd\_ac:N-Ti          & NTi           & Unseen                            &                         0      \\
    \hline
	\multicolumn{4}{|c|}{\textbf{Zr-Zn-N}} \\
    \hline
     A17B2\_hR57\_166\_cdfh\_c:Zn-Zr    & Zn17Zr2       & Unseen                            & Unseen                         \\
     AB3C3\_cF112\_227\_c\_de\_f:N-Zn-Zr & NZn3Zr3       & 0.0                               & 0.0                            \\
     ABC3\_oC20\_63\_a\_c\_cf:N-Zn-Zr    & NZnZr3        & 0.0                               & 0.0                            \\
     A12B\_tI26\_139\_fij\_a:Zn-Zr      & Zn12Zr        & 0.0                               & 0.0                            \\
     A9B8\_tP34\_137\_afg\_fg:N-Zr      & N9Zr8         & Unseen                            & Unseen                         \\
     A4B5\_tI18\_87\_h\_ah:N-Zr         & N4Zr5         & 0.34263                           & 0.34263                        \\
     A3B\_hP24\_194\_hk\_cf:Zn-Zr       & Zn3Zr         & Unseen                            & 0.0                            \\
     A3B4\_oP14\_58\_ce\_abf:N-Zr       & N3Zr4         & Unseen                            & 0.03197                        \\
     A6B5\_mC22\_12\_ij\_agh:N-Zr       & N6Zr5         & 0.0                               & 0.0                            \\
     A5B6\_mC22\_12\_agh\_ij:N-Zr       & N5Zr6         & 0.0                               & 0.0                            \\
     A2B\_cF24\_227\_c\_b:Zn-Zr         & Zn2Zr         & 0.0                               & 0.0                            \\
     AB2\_tI6\_139\_a\_e:Zn-Zr          & ZnZr2         & 0.01207                           & 0.01116                        \\
     AB\_tP4\_123\_e\_ac:Zn-Zr          & ZnZr          & Unseen                            & 0.0                            \\
     AB2\_tP6\_136\_a\_f:N-Zr           & NZr2          & 0.0                               & 0.0                            \\
     AB\_cP8\_221\_ac\_bd:N-Zr          & NZr           & Unseen                            & 0.0                            \\
    \hline
	\end{tabular}
	\label{tab:hztnzn_table}
\end{table}

\clearpage

\subsection*{Supplementary note 3: Detailed results from the massive-scale screening}

\begin{figure}[h!]
	\centering
	\includegraphics[width=1.0\textwidth]{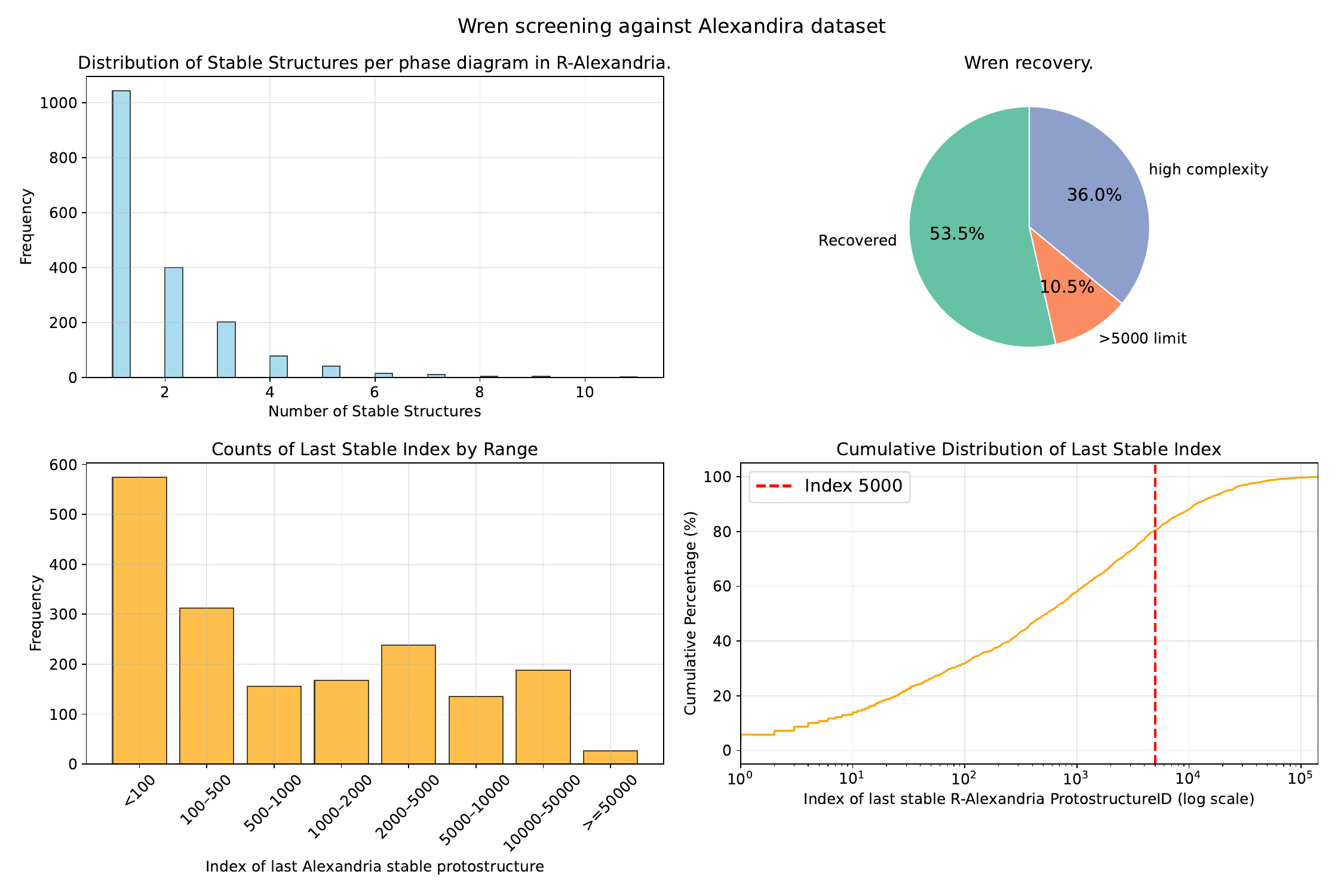} 
	\caption{\textbf{Recovery of stable phases from the Alexandria dataset using Wren screening.} 
		The plots show the cumulative fraction of Alexandria-stable phases (with complexity $\leq 5$ and $\leq 64$ atoms per cell) recovered as a function of the number of Wren-predicted candidates screened per system, ranked by energy above the Wren convex hull. 
		With a screening budget of 5,000 candidates per system, 83.6\% of the restricted-Alexandria stable phases are recovered.
	}
	\label{fig:cumulative_recovery} 
\end{figure}

Figure~\ref{fig:cumulative_recovery} shows additional details of the massive-scale screening of the paper.

\clearpage

\subsection*{Supplementary note 4: Details on fit of atomic radii}

\begin{figure*}[h!]
	\centering
	\includegraphics[width=1.0\textwidth]{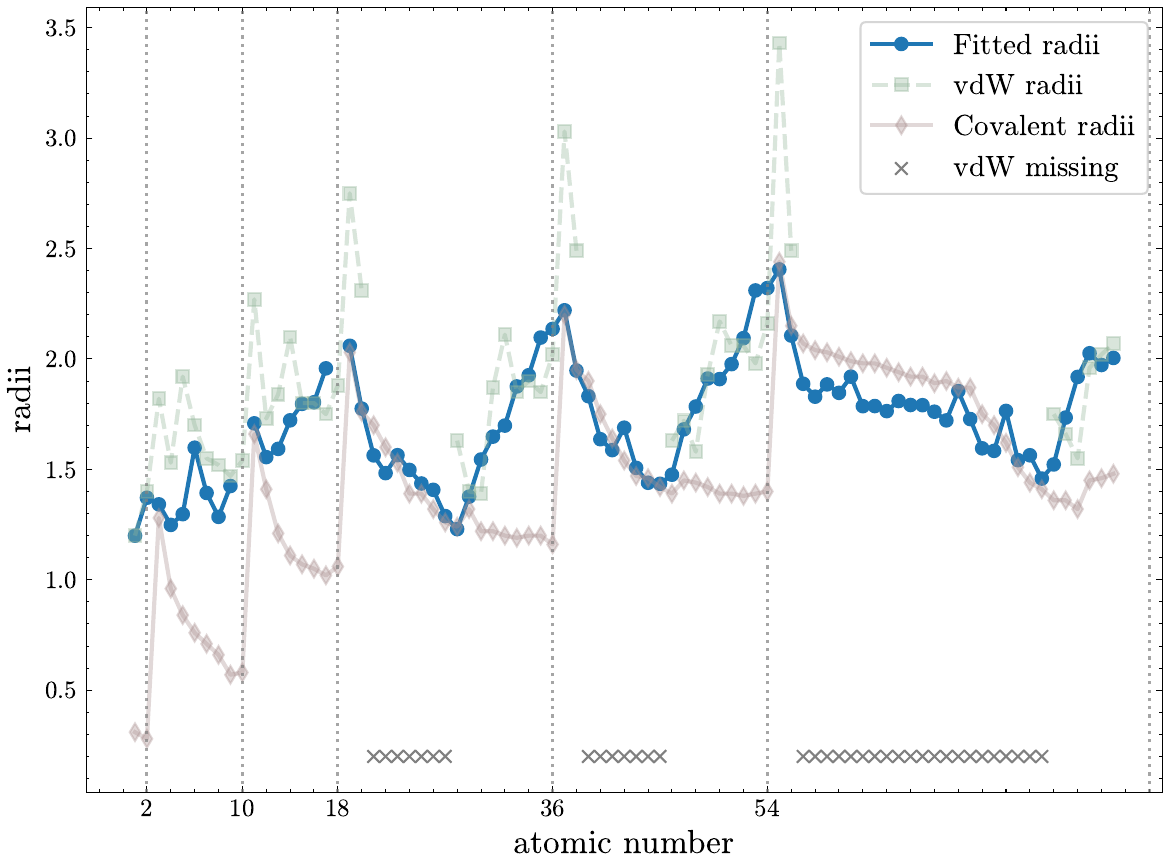}
	\caption{Comparison of fitted effective atomic radii with standard covalent and van der Waals radii. The fitted radii, derived from Materials Project database structures, provide more accurate volume estimates for crystal structure prediction by accounting for chemical environment and coordination effects.}
	\label{fig:radii_comparison}
\end{figure*}

Figure~\ref{fig:radii_comparison} shows additional details of how atomic radii are fitted for our algorithm of volume estimates.

\clearpage

\subsection*{Supplementary note 5: Novel prototypes}

Our workflow discovered 88,498 novel structural prototypes not present in either the Alexandria or GNoMe databases. A prototype is defined as a Wyckoff assignment without element specification, i.e., the symmetry template that defines a crystal structure independent of chemical composition. The large number of unique prototypes, compared to the relatively small overlap with Alexandria (1,856 shared) and GNoMe (108 shared), indicates that systematic enumeration of Wyckoff assignments accesses different regions of structure space that otherwise would not have been accessible by approaches based on substitution.
From our discovered structures, we include only prototypes with energies within 100 meV/atom of the Alexandria convex hull, ensuring these are energetically plausible templates that could host stable compounds under appropriate chemical substitution.

\begin{figure*}[h!]
\centering
\includegraphics[width=0.6\textwidth]{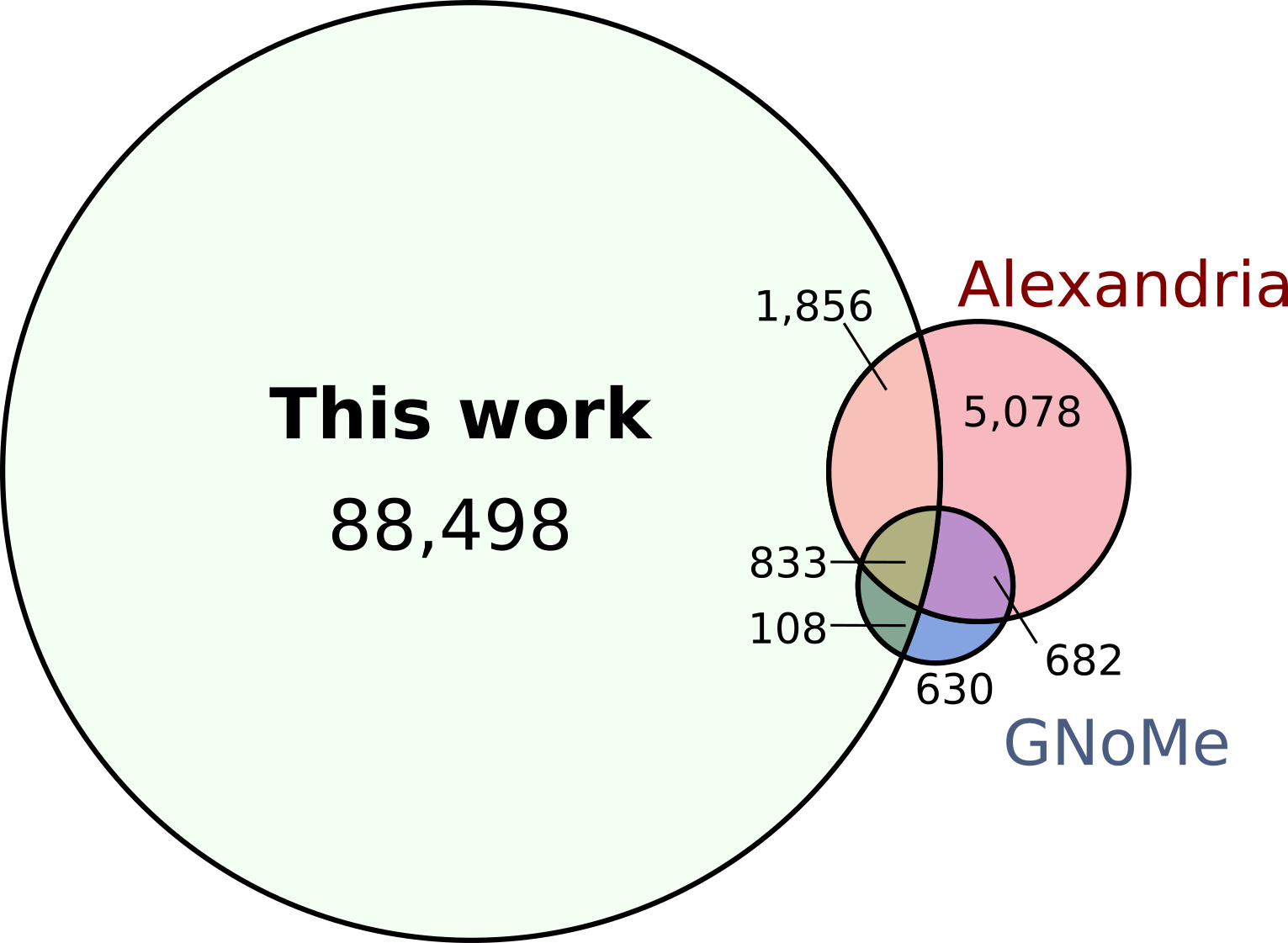}
\caption{Comparison of structural prototypes across this work, Alexandria, and GNoMe.
To ensure a meaningful comparison, we apply consistent constraints across all datasets: structures are restricted to at most 5 Wyckoff positions and 64 atoms per unit cell, matching the complexity bounds of our screening.
From our discovered structures, we include only prototypes with energies within 100 meV/atom of the Alexandria convex hull, representing candidates with realistic prospects for thermodynamic stability.
Note that we do not apply any restrictions are placed on chemical space of on Alexandria or GNoMe.
The Alexandria dataset is shown in full energy range, while for GNoMe only the stable structures are included (as only these were publicly released). The counts in this kind of Venn-like diagram gives an overdetermined equation to set the 2D disc overlaps, hence the proportionalities in the visualized areas are only approximate.}
\label{fig:novel_prototypes}
\end{figure*}

\clearpage

\subsection*{Supplementary note 6: Algorithmic details on symmetry constrained structure generation}

\begin{algorithm}[h]
\caption{Symmetry Constrained Structure Relaxation}
\begin{algorithmic}[1]
\STATE Parse protostructureID to get space group, Wyckoff positions, and lattice constraints
\STATE Sample $N$ initial DOF sets using LHC sampling $\{\mathbf{x}_i\}$, where $\mathbf{x}_i = [s_i, \mathbf{l}_i, \mathbf{w}_i]$ (volume scale, lattice DOFs, Wyckoff DOFs)
\STATE Define function $g: \mathbf{x} \mapsto (\mathcal{L}, \mathbf{R}_\mathcal{F})$ (DOF to lattice and fractional coordinates)
\STATE Define fractional-cartesian mapping  $h: \mathbf{x} \mapsto [\text{vec}(\mathcal{L}), \text{vec}(\mathbf{R}_\mathcal{F} \cdot \mathcal{L})]$

\FOR{$t = 1$ to $n_{\text{iter}}$}
    \FOR{each structure $i$}
        \STATE \textit{// Generate structure}
        \STATE $(\mathcal{L}_i, \mathbf{R}_{\mathcal{F},i}) \gets g(\mathbf{x}_i)$
        \STATE $\mathbf{R}_{\text{cart},i} \gets \mathbf{R}_{\mathcal{F},i} \cdot \mathcal{L}_i$
        
        \STATE \textit{// MLIP for energy, forces, and stress}
        \STATE $E_i, \mathbf{F}_{R,i}, \sigma_i \gets \mathrm{MLIP}(\mathcal{L}_i, \mathbf{R}_{\text{cart},i})$
        
        \STATE \textit{// Compute generalized force on lattice (Ref.~\cite{lenz2019parametrically})}
        \STATE $V_i \gets \det(\mathcal{L}_i)$
        
    \STATE $\displaystyle\mathcal{F}_{\mathcal{L},i} \gets -\mathcal{L}_i^{-T} V_i \sigma_i - \mathbf{R}_{\mathcal{F},i}^T \mathbf{F}_{R,i}$

        \STATE \textit{// Project forces onto DOFs using VJP, computed using JAX}
        \STATE $\nabla_{\mathbf{x}} E_i \gets \left(\dfrac{\partial h}{\partial \mathbf{x}_i}\right)^T [\text{vec}(\mathcal{F}_{\mathcal{L},i}); \text{vec}(\mathbf{F}_{R,i})]$
    \ENDFOR
    
    \STATE Update all $\mathbf{x}_i$ using optimizer with gradients $\nabla_{\mathbf{x}} E_i$
    \IF{$\max_i \|\nabla_{\mathbf{x}} E_i\| < 0.001$}
        \STATE \textbf{break}
    \ENDIF
\ENDFOR

\STATE \textbf{return} relaxed structures $\{(\mathcal{L}_i, \mathbf{R}_{\mathcal{F},i})\}$

\end{algorithmic}
\end{algorithm}s
\newpage
\subsection*{Supplementary note 7: Supporting data for figure 5}

The table shows the cumulative counts of protostructures with decomposition enthalpy below each threshold defined in column $\boldsymbol{E_{\mathrm{hull}}}$ \textbf{(eV/atom)}. \textbf{Alexandria} indicates structures within the screened chemical space with no other restrictions (such as complexity or number of atoms). Column \textbf{Unique} shows the count of protostructures that exist in our screened set but not in Alexandria, and have a decomposition enthalpy below 100 meV/atom on the convex hull formed by the union of our structures and Alexandria. \textbf{This work} indicates the total number of structures identified that satisfy the energy above convex hull threshold defined by structures present in this workflow.

\begin{table}[h]
\centering
\begin{tabular}{rrrr}
\hline
$\mathrm{E_{hull}}$ (eV/atom) & Alexandria & Unique & This work \\
\hline

               0    &         4407 &     2303 &        4710 \\
               0.01 &         9167 &     9522 &       13488 \\
               0.02 &        13656 &    22031 &       27132 \\
               0.03 &        18584 &    40690 &       46734 \\
               0.04 &        23763 &    66777 &       73616 \\
               0.05 &        28922 &   101772 &      109387 \\
               0.06 &        33868 &   146541 &      155188 \\
               0.07 &        38483 &   201579 &      212490 \\
               0.08 &        43011 &   265442 &      281131 \\
               0.09 &        47318 &   333179 &      361226 \\
               0.1  &        51378 &   391532 &      456110 \\
\hline
\end{tabular}
\end{table}

\end{document}

%% file: alias.tex
\definecolor{todoColor}{HTML}{E74C3C}
\definecolor{fixmeColor}{HTML}{D35400}
\definecolor{noteColor}{HTML}{2ECC71}
\definecolor{questionColor}{HTML}{8E44AD}
\definecolor{highlightColor}{HTML}{F1C40F}

\newcommand{\todo}[1]{\textcolor{todoColor}{\texttt{TODO:} #1}}
\newcommand{\fixme}[1]{\textcolor{fixmeColor}{\texttt{FIXME:} #1}}
\newcommand{\note}[1]{\textcolor{noteColor}{\texttt{NOTE:} #1}}
\newcommand{\question}[1]{\textcolor{questionColor}{\texttt{?} #1}}
\newcommand{\highlight}[1]{\colorbox{highlightColor}{#1}}

\newcommand{\protoid}[1]{\texttt{\detokenize{#1}}}

\newcommand{\symgen}{\texttt{httk\_symgen}}
\newcommand{\umlips}{uMLIPs\ }

\newcommand{\macempa}{\texttt{MACE\_MPA-0}}
\newcommand{\orb}{\texttt{ORB v2}}

\newcommand{\chem}[1]{\ce{#1}}